\documentclass[fleqn,twoside]{article}
\usepackage{espcrc2,html}

\usepackage{graphicx}

\newcommand{\AmS}{{\protect\the\textfont2
  A\kern-.1667em\lower.5ex\hbox{M}\kern-.125emS}}

\newcommand{\uno}{\one}
\newcommand{\one}{{\mathbf 1}}
\newcommand{\sign}[1]{{{\rm \,sign(} #1 {\rm )}}}
\newcommand{\eq}[1]{eq.~(\ref{#1})}

\usepackage[figuresright]{rotating}

\title{A comparative study of numerical methods for the overlap Dirac
operator---a status report}

\author{
\vspace{-4cm}
\noindent\hspace*{\fill}BUGHW-SC 2001/6\\
\noindent\hspace*{\fill}WUB 01-12\\ 
\vspace{3cm}
J.\ van den Eshof\thanks{Partly supported by EC grant
       HPRN-CT-2000-00145 Hadrons/Lattice QCD.}\address[Utrecht]{Department of Mathematics, 
                     University of Utrecht, The Netherlands},
A.\ Frommer\address[MathWup]{Department of Mathematics, 
                     University of Wuppertal, Germany},
Th.\ Lippert\address[PhysWup]{Department of Physics, 
                     University of Wuppertal, Germany},
K.\ Schilling\addressmark[PhysWup] and
H.\ van der Vorst\addressmark[Utrecht]}

\begin{document}

\begin{abstract}
  Improvements of various methods to compute the sign function of the
  hermitian Wilson-Dirac matrix within the overlap operator are
  presented.  An optimal partial fraction expansion (PFE) based on a
  theorem of Zolotarev is given.  Benchmarks show that this PFE
  together with removal of converged systems within a multi-shift CG
  appears to approximate the sign function times a vector most
  efficiently.  A posteriori error bounds are given.
\end{abstract}

\thispagestyle{empty}\maketitle

\section{INTRODUCTION}

The overlap operator, $D = \uno + r\gamma_5\sign{Q}$, satisfies the
Ginsparg-Wilson relation and thus exhibits chiral symmetry at finite
lattice spacing $a$ (see \cite{Neu00} and references therein).
However, due to the sign function of the hermitian Wilson-Dirac
operator, $Q$, its numerical evaluation is extremely costly, with an
overhead estimated to be at least a factor ${\cal O}(100)$
compared to Wilson fermions.

In this status report of our ongoing interdisciplinary project, we
demonstrate that well established methods to compute the sign function
like Lanczos and multi-shift CG in combination with a partial fraction
expansion (PFE/CG) can be improved substantially.  We present
benchmarks of Neuberger's PFE/CG method \cite{Neu00}, an optimal
PFE/CG method with reduced number of poles, a PFE-improved version of
Bori\c ci's Lanczos process for $Q^2$ \cite{Bor99c}, as well as the
standard Chebyshev approximation. It turns out that the PFE/CG method
with removal of converged systems is most efficient.

Furthermore, for error monitoring and termination of iterations, we
derive {\em a posteriori} error bounds of the approximation of the
sign function for both Lanczos (in terms of the residual of a related
CG-process) and PFE methods (in terms of the residuals in the
multi-shift solver).

\section{NUMERICAL PROBLEMS}

Computations involving the overlap operator, $D$,
are characterized by two nested iterations, {\em (i)}
the 
outer iterative solution of 
\begin{equation}
\label{FIRST}  
D\, x = (\uno + r\gamma_5\sign{Q})\, x = b\qquad |r| \le 1,
\end{equation}
requiring {\em (ii)} an inner iteration for $s$
\begin{equation}
s=\sign{Q}b
\end{equation} 
in each outer iteration step.

Despite the fact that nested schemes are sub-optimal as information
built-up for the sign function is discarded after each iteration, they
might still be superior to alternatives from \cite{Neu00}.

\section{NUMERICAL METHODS FOR $S\, x$\label{METHODS}}

\subsection{Polynomial approximations for $t^{-\frac{1}{2}}$}

These methods determine polynomials $p_k$ which approximate
$t^{-\frac{1}{2}}$ on $[a^2,b^2]$ with $a\le |\lambda_{\mbox{\tiny
    min}}|$ and $b\ge |\lambda_{\mbox{\tiny max}}|$, the extremal
eigenvalues of $Q$. The approximation to $s=\sign{Q}b$ is then
$s\approx Q\,p_k(Q^2)\,b$. Polynomials that have been used are {\em Chebyshev
  polynomials} with linear convergence (error $\propto(\frac{\kappa-1}{\kappa+1})^{k}$,
$\kappa$ being the condition number of $Q$), {\em Legendre
  polynomials} as applied in Ref.~\cite{Hernandez:1998et}, {\em
  Gegenbauer polynomials} as introduced by Bunk
(Ref.~\cite{Bunk:1998wj}) and {\em Schulz polynomials} (error $\propto
(\frac{\kappa-1}{\kappa+1})^{k^{\frac{2}{3}}}$) which will be presented in a forthcoming
publication of our collaboration.

\subsection{Lanczos based methods}
The Lanczos process in matrix form reads
\begin{equation}
Q\, V_k = V_k T_k +\beta_{k+1} v_{k+1} e_k^T,\; \mbox{with}\;
V_k^{\dagger}b =e_1,
\label{LANCZOSONE}
\end{equation} where we assume $\|b\|=1$.
We refer to \eq{LANCZOSONE} as ``Lanczos for $Q$''.  Two ways have
been proposed to approximate $\sign{Q}b$ by diagonalization of $T_k$:
\begin{equation}
\sign{Q}b \approx Q\, V_k\left( T_k^2 \right)^{-\frac{1}{2}}e_1. 
\quad\mbox{(\cite{Bor99a})}
\end{equation}

\begin{equation}
\sign{Q}b \approx V_k\sign{T_k}e_1. \quad\mbox{(\cite{Vor00})}  
\end{equation}
The errors of both methods are highly oscillating as a function of
$k$.  For the second one, the peaks are bounded, however.  In order to
avoid such oscillations Bori\c ci has introduced an alternative based
on a Lanczos process on $Q^2$ \cite{Bor99c}:
\begin{equation}
Q^2\,V_k = V_k T_k +\beta_{k+1} v_{k+1} e_k^T, \; \mbox{with}\;
V_k^{\dagger}b=e_1.
\end{equation}
\begin{equation}
\sign{Q}b \approx Q\, V_kT_k^{-\frac{1}{2}}e_1.
\end{equation}
The latter method (``Lanczos for $Q^2$'') shows a smoother convergence
rate as well as a potentially smaller projected system $T_k$.
However, in any case the spectral decomposition of $T_k$ is
computationally very costly.

\subsection{Partial Fraction Expansion and multi-shift CG (PFE/CG)}
The elegant idea to use a fixed number of vectors by means of partial
fractions expansions has been proposed by Neuberger
\cite{Neu00}:
\begin{eqnarray}
\sign{Q}b\! \approx\! x^{\mbox{\tiny PFE}}\!\!\!\!=\!\!
\sum_{i=1}^m  \frac{\omega_i\,Q}{Q^2\!\!+\!\!\tau_i}b
           \! \approx\! x_k\! =\!\! \sum_{i=1}^m \omega_i Q x_k^i.\!\!\!\!\!
\end{eqnarray}
The $m$ vectors $x_k^i$ are computed in step $k$ of the multi-shift CG
method \cite{Glassner:1996gz} for the shifts $\tau_i$.

Two rational approximations so far have been applied in the context of
the overlap operator:

\noindent{\bf Neuberger's proposal} (\cite{Neu00}): The
  coefficients are defined  by 
\begin{equation}
\begin{array}{l}
\tau_i=\tan^2\left( \frac{\pi}{2m}(i-\frac{1}{2})\right)\\
\omega_i=\frac{1}{m}\cos^{-2}\!\left( \frac{\pi}{2m}(i-\frac{1}{2})\right).\\
\end{array}
\end{equation}
In general, a large number $m$ of poles $\tau_i$ is required to achieve
practical precisions.

\noindent{\bf Remez algorithm} (Edwards et al. \cite{EHN98}): By use of the Remez
algorithm, an optimal approximation $g(x)$ to $x^{-\frac{1}{2}}$ in
$\|\cdot\|_{\infty}^{[\lambda_{\mbox{\tiny
      min}}^2,\lambda_{\mbox{\tiny max}}^2]}$ is constructed,
resulting in a substantially smaller number of poles.  However, the sign
function is aproximated as $x\,g(x^2)$ which is {\em not} the
$\|\cdot\|_{\infty}$-optimal approximation to $\sign{x}$ in
$[-|\lambda_{\mbox{\tiny max}}|,-|\lambda_{\mbox{\tiny min}}|]\cup
[|\lambda_{\mbox{\tiny min}}|,|\lambda_{\mbox{\tiny max}}|]$.

\section{IMPROVEMENTS}

\subsection{Lanczos procedures}

As mentioned, the Lanczos approach might be slow since a
diagonalization of $T_k$ is required. As far as the ``Lanczos on
$Q^2$'' approach is concerned we propose to use a PFE, as detailed
next, to compute a first approximation to the inverse square root of
the {\em full matrix} $(T_k)$. Based on this approximation, the
Lanczos procedure is repeated to yield the final approximation to
$(Q^2)^{-\frac{1}{2}}b$.


\subsection{PFE/CG}
The vector updates in PFE/CG  play a significant role for large
numbers of poles for practical implementations. Therefore, we seek for
a reduction of the number of poles to improve PFE/CG. In contrast to
Ref.~\cite{EHN98} we try to find a rational function $f(x)$ that
minimizes 
\begin{equation}
|| 1-\sqrt{x}f(x)||_{\infty}^{[\lambda_{\mbox{\tiny
    min}}^2,\lambda_{\mbox{\tiny max}}^2]}.  
\end{equation}
Then $x\,f(x^2)$ is the $\|\cdot\|_{\infty}$-optimal rational
approximation of the sign function on $[-|\lambda_{\mbox{\tiny
    max}}|,-|\lambda_{\mbox{\tiny min}}|] \cup [|\lambda_{\mbox{\tiny
    min}}|,|\lambda_{\mbox{\tiny max}}|]$.
By means of Zolotarev's theorem
\cite{IDK00} $f(x)$ can be given in analytic form:
\begin{equation}
f(x) = D\frac{\prod_{l=1}^{k-1}
(x+c_{2l})
}{\prod_{l=1}^{k}
(x+c_{2l-1})},  
\end{equation}
where the coefficients can be expressed in terms of Jacobian
elliptic functions:
\begin{equation}
c_l =
\frac{\mbox{sn}^2(lK/2k;\kappa)}{1-\mbox{sn}^2(lK/2k;\kappa)},\;l=1,\dots,2k-1,  
\end{equation}
with 
$\sqrt{1-\kappa^2}=
|\frac{\lambda_{\mbox{\tiny min}}}{\lambda_{\mbox{\tiny max}}}|$ and $D$
being uniquely determined by the condition
\[
\max_{C[1,(\frac{\lambda_{\mbox{\tiny max}}}{\lambda_{\mbox{\tiny
        min}}})^2]}[1-\sqrt{x}f(x)]
=-\!\!\!\!\!\!\!\! 
\min_{C[1,(\frac{\lambda_{\mbox{\tiny max}}}{\lambda_{\mbox{\tiny
        min}}})^2]}[1-\sqrt{x}f(x)].\]
Table 1 shows that the method drastically reduces the number of
poles, in particular for large condition numbers.
\begin{table}[!htb]
\vspace*{-12pt}
\caption{Number of poles for precision 0.01.}
\label{table:1}
\renewcommand{\tabcolsep}{10pt} 
\renewcommand{\arraystretch}{.9} 
\begin{tabular}{@{}l|ccc}
\hline
$\frac{\lambda_{\mbox{\tiny max}}}{\lambda_{\mbox{\tiny min}}}$ &
Neuberger & \mbox{Remez} \cite{EHN98} & \mbox{Zolotarev}\\
\hline
200 & 19 & 7 & 5\\[.1cm]
1000 & 42 & 12 & 6\\[.1cm]
100000 & $> 500$ & ? & 10\\[.1cm]
\hline
\end{tabular}
\vspace*{-12pt}
\end{table}

Another interesting idea is to remove converged systems from the multi-shift
process early, as residuals for shifted matrices with large shifts $\tau_i$
reduce more quickly.  Under some restrictions on the quality of PFE
we can show that for a total error of at most $\epsilon$ we can stop
updating system $j$ after step $k$ as soon as
\begin{equation}
||r_k^j|| \le \frac{\epsilon}{m}\frac{\sqrt{\tau_j}}{\omega_j}.
\end{equation}

\section{A POSTERIORI ERROR BOUNDS}

The error in the PFE/CG method is composed of 2 parts.
For a total error of at most $\epsilon$ we demand:
\begin{equation}
\mbox{I.}\; 
|\sign{\lambda} - \sum_{i=1}^m 
\omega_i \frac{\lambda}{\lambda^2+\tau_i}|\le {\epsilon}/2,
\end{equation}
\begin{equation}
\mbox{II.}\; || x^{\mbox{\tiny PFE}} -x_k || \le {\epsilon}/2.  
\end{equation}
One can prove that the total error $\le\epsilon$, if the CG residual
for the smallest shift satisfies
\begin{equation}
||r_k^1|| \le \frac{\epsilon}{2+\epsilon}.  
\end{equation}

For ``Lanczos for $Q^2$'' it is worth noting that 
we also managed to get a posteriori error bounds which will be
presented in a forthcoming paper.

\section{NUMERICAL EXPERIMENTS}

Our tests have been carried out on quenched $16^4$ configurations at
$\beta=6.0$ and $m=-1.6$ with the error for the approximation of the
sign function being $<10^{-10}$.  The timings are from 16 nodes of the
Wuppertal cluster computer ALiCE.

\begin{table}[!tb]
\caption{Benchmarks.}
\label{table:2}
\small\renewcommand{\arraystretch}{.85} 
\begin{tabular}{@{}l|lllll}
\hline
confs & 1 & 2 & 3 & 4 & 5\\
\hline
\hline
$|\lambda_{\mbox{\tiny min}}|\cdot 10^{3}$ & $0.455$ & $1.39$ & 
$1.17$ & $2.23$ & $3.02$ \\
$|\lambda_{\mbox{\tiny max}}| $ & $2.48$ & $2.48$ & 
$2.48$ & $2.48$ & $2.48$ \\
poles Neub. & 143 & 82 & 89 & 65 & 56 \\
poles Zolo. & 21 & 18 & 19 & 17 & 16 \\
\hline
\multicolumn{6}{c}{Chebyshev}\\
\hline
MVs & 9501 & 3501 & 4001 & 2301 & 2201 \\
time/s & 655 & 247 & 278 & 160 & 154 \\
\hline
\multicolumn{6}{c}{Lanczos/PFE}\\
\hline
MVs & 2281 & 1969 & 1953 & 1853 & 1769 \\
time/s & 150 & 131 & 129 & 124 & 118 \\
\hline
\multicolumn{6}{c}{PFE/CG Neuberger}\\
\hline
MVs & ? & 985 & 977 & 929 & 887 \\
time/s & ? & 340 & 362 & 274 & 215 \\
\hline
\multicolumn{6}{c}{PFE/CG Zolotarev without removal}\\
\hline
MVs        & 1141 & 985  & 977 & 927 & 885 \\
 time/s & 154  & 125  & 125 & 116 & 102 \\
\hline
\multicolumn{6}{c}{PFE/CG Zolotarev $+$ removal}\\
\hline
MVs & 1205 & 1033 & 1033 & 971 & 927 \\
time/s & 122 & 93 & 97 & 87 & 79 \\
\hline
\end{tabular}
\vspace*{-10pt}
\end{table}

\section{OUTLOOK}
\vspace*{-.1cm}
The benchmark results demonstrate that the PFE/CG/Zolotarev procedure
with removing of converged systems turns out to be most effective.

As a next step, we will tune the accuracy of the sign approximation
within the solution of the outer problem, $D\,x=b$ (\eq{FIRST}).
Furthermore, we will investigate the effect of projecting out some low
eigenvalues of $Q$ onto our findings.

\end{document}